\documentclass[aps,showpacsx,twocolumn]{revtex4-1}
\usepackage{float}
\usepackage{graphicx}
\usepackage{caption}
\usepackage{amsmath}
\usepackage{array}
\usepackage{bm}
\usepackage{amssymb}
\usepackage{amsfonts}
\usepackage{amssymb}
\usepackage{color}

\begin{document}
\title{pion-rho mixing as a mechanism for non-monotonic charged pion behavior in magnetic fields}
\author{Ziyue Wang}
\affiliation{School of Physics and Optoelectronic Engineering, Beijing University of Technology, Beijing 100021, China}
\date{\today}
\begin{abstract}
We investigate whether magnetic field induced $\pi-\rho$ mixing can explain the non-monotonic behavior of the charged pion reported in lattice QCD. Using a near-pole effective action derived from the SU(2)$_f$ Nambu--Jona-Lasinio model, we show that the lowest Landau level charged pion mixes with the longitudinally polarized charged rho meson, which shares the same quantum numbers in a magnetic background. The resulting level repulsion is strongly amplified by the suppression of the rho wave-function renormalization near the pole. As a consequence, the lowest mixed mode develops a turnover as the magnetic field increases, reproducing the qualitative trend seen on the lattice. Comparison with the direct determinant solution of the Landau-projected kernel shows that the mechanism is robust, although the quantitative location of the maximum remains scheme dependent. These results support $\pi-\rho$ mixing as an important candidate mechanism for charged meson spectra in strong magnetic fields.
\end{abstract}
\maketitle

\section{Introduction}
Strong magnetic fields, reaching up to $10^{19}$ Gauss ($|eB|\sim 10 m_\pi^2$) realized in a variety of physical environments, including the early universe, magnetars, and non-central heavy-ion collisions~\cite{Grasso:2000wj, Durrer:2013pga, Kiuchi:2015sga, Skokov:2009qp, Deng:2012pc} can significantly influence the behavior of QCD matter~\cite{Kharzeev:2007jp, Kharzeev:2010gr, Gursoy:2014aka, Gusynin:1995nb, Kharzeev:2010gd, Bali:2012zg, Tomiya:2019nym, Bali:2014kia, Bali:2012jv}. Comprehensive reviews of QCD in strong magnetic fields can be found in Refs.~\cite{Miransky:2015ava, Fukushima:2012kc, Andersen:2014xxa}. At such intensities, magnetic field induced QED effects become comparable to hadronic scales and can qualitatively modify both the phase structure of QCD matter and the properties of QCD bound states. In particular, the spectrum of charged mesons is strongly affected: their transverse motion is quantized into Landau levels, while their internal structure can be significantly distorted by the magnetic field.

Recent lattice QCD simulations have focused on the QCD properties in magnetic field~\cite{Bali:2011qj, Bali:2012zg, Ilgenfritz:2013ara}  and revealed an unexpected feature of charged mesons  \cite{Ding:2020hxw, Ding:2020jui, Ding:2026qzu}: the effective mass of the charged pion exhibits a non-monotonic dependence on the magnetic field strength. While the mass initially increases at weak fields, consistent with the lowest Landau level expectation, it reaches a maximum at intermediate fields and subsequently decreases as the field is further increased. This behavior cannot be explained by Landau quantization alone and has remained a long-standing discrepancy between lattice results and model descriptions.

Light meson properties in magnetic fields have been intensively investigated in effective models and continuum approaches ~\cite{Inagaki:2003yi, Yu:2014xoa, Fayazbakhsh:2014mca, Coppola:2018vkw, Coppola:2019uyr, Chaudhuri:2019lbw, Kamikado:2013pya, Mamo:2015dea, Li:2016gfn}, including NJL-type models, functional methods, and holographic descriptions. Vector mesons in magnetic backgrounds, in particular the charged $\rho$, have also been studied extensively \cite{Liu:2014uwa, Ghosh:2016evc, Kawaguchi:2015gpt}, where modifications of polarization structure and possible condensation phenomena were discussed. More recent studies have attempted to explain the lattice behavior by incorporating magnetic field dependent quark couplings \cite{Avancini:2021pmi}, meson mixing effects \cite{Carlomagno:2022arc, Coppola:2023mmq}, wave-function renormalization \cite{Wen:2023qcz}, 
or inverse magnetic catalysis \cite{Li:2023rsy}. However, none of these approaches has been able to reproduce the lattice trend in a fully satisfactory manner, suggesting that an additional dynamical mechanism becomes operative at strong magnetic fields. 

In this Letter, we investigate $\pi^\pm-\rho^\pm_{s_z=0}$ mixing as a possible mechanism for the non-monotonic magnetic field dependence of the charged pion mass seen in lattice QCD. In a magnetic background, these two modes share the same conserved quantum numbers and can therefore mix. Using a near-pole effective action derived from the SU(2)$_f$ Nambu--Jona-Lasinio model, we show that the associated level repulsion is strongly amplified by the suppression of the longitudinal-rho residue near the pole. The resulting lowest mixed mode develops a turnover with increasing magnetic field. We further compare this near-pole description with the direct determinant solution of the Landau projected kernel, which supports the same qualitative mechanism while indicating quantitative sensitivity to the detailed structure of the kernel. The present Letter focuses on the mechanism itself; a more systematic comparison between different extraction strategies will be presented elsewhere.

This Letter is organized as follows. In Sec.~\ref{toy}, we first illustrate the basic physical picture of magnetic field induced $\pi-\rho$ mixing through a simple toy-model analysis. In Sec.~\ref{NJLcalculation}, we then formulate the field theoretical description based on the Landau-projected NJL kernel. In Sec.~\ref{result}, we present the numerical results for the unmixed modes, wave-function renormalizations, mixing strength, and the resulting lowest eigenmode. Finally, Sec.~\ref{summary} contains our summary and discussion.

\section{A toy-model illustration of $\pi-\rho$ mixing}
\label{toy}
To clarify the idea of $\pi-\rho$ mixing in magnetic field, we first look at a similar system, and then analyze $\pi-\rho$ system in a quantum mechanical toy model. Positronium is the bound state of an electron and a positron, it exists in two quantum spin states singlet (para-positronium) and triplet (ortho-positronium). In a magnetic field $\vec{B}$, the interaction with the magnetic moments of the electron and positron lifts the degeneracy of the triplet states. Crucially, the $m_s = 0$ component of the triplet and the singlet (S=0) state are no longer pure eigenstates of the Hamiltonian. Both states share the same projection of total spin along the magnetic field $m_s = 0$. The magnetic interaction, proportional to $\vec{B}\cdot(\vec{\mu}_{e^-} + \vec{\mu}_{e^+})$, couples these two states because their magnetic quantum numbers are identical. This coupling causes them to mix, forming two new energy eigenstates that are superpositions of the original singlet and triplet $m_s=0$ states. This mixing alters their properties. Most notably, the originally long-lived ortho-positronium $(m_s=0)$ acquires a small admixture of the short-lived para-positronium. This gives it a new, faster decay channel into $\gamma\gamma$, reducing its observable lifetime. This is a direct quantum mechanical analogue to the proposed $\pi-\rho$ mixing in QCD under a strong magnetic field, where the pion (pseudoscalar, like the singlet) and the longitudinally polarized rho meson (vector, like the $m_s=0$ triplet) can mix because they share the same quantum numbers in the magnetic field.

\begin{figure}[H]
\centering
\includegraphics[width=0.4\textwidth]{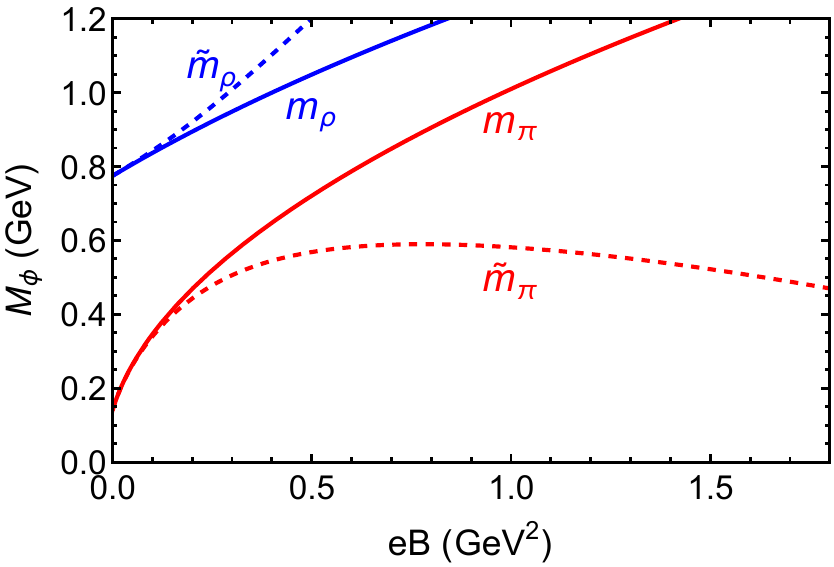}
\caption{The mass eigenvalues of the $\pi-\rho$ mixing state (dashed lines) in comparison with the mass of unmixed state (solid line).}
\label{fig0}
\end{figure}

We first give a simple picture of how this mixing would change the charged meson mass. Consider a meson composed by $u\bar{d}$ quark. The most light system are the pseudo-scalar $\pi^+$ ($J^P=0^-$, 135MeV), vector meson $\rho^+$ ($J^P=1^-$, 770MeV). In the external magnetic field along z-direction, there are additional interaction between magnetic moment of the constituent quark and the magnetic field 
\begin{eqnarray}
\hat{H}_B=-(\vec{\mu}_u+\vec{\mu}_{\bar{d}})\cdot \vec{B}=-\frac{eB}{3m}(2S_{uz}+S_{dz})
\end{eqnarray}
The spin state of the constituent quark longitudinally polarized rho is $ |1\rangle=(|\uparrow\downarrow\rangle+|\downarrow\uparrow\rangle)/\sqrt{2}$ and for pion $|2\rangle=(|\uparrow\downarrow\rangle-|\downarrow\uparrow\rangle)/\sqrt{2}$, where in $|\uparrow\downarrow\rangle$ the former one is spin of $u$-quark and the latter one is spin of $\bar{d}$-quark. In the subspace of $(\rho^+_{s_z=0},\pi^+)$, the non-vanishing components $\hat{H}_B$ of are $\langle 1|\hat{H}_B|2\rangle=\langle 2|\hat{H}_B|1\rangle=-\frac{eB}{6m}$ in natural unit, with $m$ the mass of constituent quark. Taking also meson mass and lowest Landau level  into consideration, the Hamiltonian in $(\rho^+_{s_z=0},\pi^+)$ subspace becomes
\begin{equation}
\left(\begin{array}{cc}
\sqrt{m_\rho^2+eB}   &  -\frac{eB}{6m}  \\
-\frac{eB}{6m}    &  \sqrt{m_\pi^2+eB}    \\
\end{array}\right). 
\end{equation}
The eigenvalues of the Hamiltonian $\tilde{m}_\pi$ and $\tilde{m}_\rho$ stand for the mixed meson mass in magnetic field. With $m=0.308$GeV, as shown in Fig.\ref{fig0}, one can explicitly find the lower band $\tilde{m}_\pi$ becomes non-monotonic in magnetic field, while the unmixed pion and rho becomes heavier due to the Landau quantization. This down-bending of the lower mode originates from the level repulsion of the mixing effect. 
The above description gives a quantum mechanical picture of the mixing and the resulting level repulsion. However, it can only be treated as a toy model to illustrate the idea, a quantum‑mechanical treatment for light mesons such as the pion is by nature inadequate, and a more rigorous field‑theoretical framework is required. In the next section, we perform a field‑theoretical treatment based on near-pole expansion starting from the NJL model. 

\section{Field Theoretical calculation}
\label{NJLcalculation}
In order to systematically describe the meson mass in magnetic field, we adopt the SU(2)$_f$ Nambu–Jona-Lasinio model with quarks minimally coupled to the background field $\mathcal{L}=\bar{\psi}(iD\!\!\!\!/-m_0)\psi+G_S[(\bar{\psi}\psi)^2+(\bar{\psi}i\gamma^5\psi)^2]-G_V[(\bar{\psi}\gamma^\mu\tau^a\psi)^2]$, with $\tau^a$ representing the isospin Pauli matrices, $G_S$ and $G_V$ the coupling constants with respect to the scalar (pseudoscalar) and the vector channels \cite{Klevansky:1992qe, Hatsuda:1994pi, Vogl:1991qt}. The covariant derivative couples quarks to an external magnetic field $\vec{B}=B\hat z$ along the positive $z$ direction. For the quark propagator we use Landau-level representations \cite{Schwinger:1951nm, Miransky:2015ava}. 

In the QCD vacuum, quark-antiquark pairs condense, leading to the generation of dynamical masses for quarks. Assuming isospin symmetry with $m_u = m_d = m_0$, the constituent quark mass $m_f$ for light quarks is determined by solving the gap equation $m_f~=~m_0+2G_S\text{Tr}S(x,x)$. The gap equation at zero temperature is 
\begin{eqnarray}
\label{gap}
m_f=m_0+8 G_S m_f N_c\sum_{f=u,d}\frac{|q_fB|}{4\pi}\sum_{n=0}^{\infty} \alpha_n\int_{-\infty}^\infty\frac{dp_3}{2\pi}\frac{1}{2E_f},
\end{eqnarray}
where $\alpha_n=1 $ for $n=0$, and $\alpha_n=2 $ for $n\geq1$ is the degree of degenerency of spin. $n_F=1/(e^{x/T}+1)$ is the Fermi Dirac distribution, the energy eigenvalue of quark in the magnetic field $E_f~=~\sqrt{2n|q_fB|+p_3^2+m_f^2}$. This NJL mean field gap equation reproduces the magnetic catalysis at zero temperature as expected in other model analysis.

\subsection{Landau-projected polarization function}
After Hubbard-Stratonovich transformation, integrating out the quark fields, and expanding to quadratic order, one arrives at a matrix of polarization functions for the charged pion and rho channels, 
\begin{eqnarray}
\label{quadraaction}
S^{(2)}_{\rm eff}
&=&
\frac{1}{2} \int d^4x d^4y
\Phi^\dagger(x)
[
\mathcal{G}^{-1}\delta(x-y)-\Pi(x,y)
]
\Phi(y)\nonumber\\
&&
+S^{(2)}_{\rm tree},
\end{eqnarray}
where $[\mathcal{G}^{-1}\delta(x-y)-\Pi(x,y)]$ is the kernel in the reduced subspace of $\pi^+$ and $\rho^+_{s_z=0}$ denoted by $\Phi=(\pi,\rho_3)^T$, $\mathcal G^{-1}=\text{diag}(1/2G_S,1/2G_V)$ is the contact inverse propagator, and $\Pi$ is the polarization matrix from the quark loop.  $S^{(2)}_{\rm tree}=\int d^4x\mathcal L_{\text{tree}}$ comes from the tree level counterterm, which will be determined in Sec.\ref{matching}. In the Random Phase Approximation (RPA), the core ingredient is the polarization function, which is defined here by 
\begin{eqnarray}
\Pi(x,y)&=&\left(\begin{array}{cc}
\Pi_{\pi\pi} &  \Pi_{\pi\rho}^3 \\
\Pi_{\rho\pi}^3  &  \Pi_{\rho\rho}^{33} \\
\end{array}\right).
\end{eqnarray}
In coordinate space, these polarization functions are defined by $\Pi_{ab}=-i\text{Tr}\big[\Gamma_a S_f(r,r')\Gamma_bS_f(r',r)\big]$. The diagonal components are the ordinary polarization of charged pion $\Gamma_{a,b}=(i\gamma^5\tau^\mp)$ and $\rho_3^\pm$ with $\Gamma_{a,b}=(\gamma^3\tau^\mp)$. The off-diagonal component $\Pi_{\rho\pi}^3$ and $\Pi_{\pi\rho}^3$ are non-vanishing in presence of the external magnetic field, with $\Gamma_a=(\gamma^3\tau^-)$ and $\Gamma_b=(i\gamma^5\tau^+)$ for the former one and the exchange for the latter. They introduce the mixing between $\pi^\pm$ and $\rho^\pm_{s_z=0}$ at loop level.
 
To focus ourselves on the given Landau level modes in the bosonic action, we expand the meson field by their eigenfunctions in magnetic field 
\begin{eqnarray}
\phi(x)=\sum_{n,q_y,q_z}\phi_n(q_0,q_3)\varphi_{n}(x_\perp)e^{-iq_0t+iq_3z},
\end{eqnarray}
with $\phi$ stands for $\pi^+$ or the longitudinal $\rho^+$, and $\varphi_{n}$ is the standard gauge dependent Landau wavefunctions 
\begin{equation}
\begin{split}
\label{wavefunction}
\varphi_{n}(r)&=e^{-ip_0 t+ip_3z}\varphi_{n,p_1}(\vec{r}_\perp)\nonumber\\
&=e^{-ip_0 t+ip_3z}\frac{e^{-\frac{\left(\frac{y}{l}+s_\perp p_1 l\right)^2}{2}}}{\sqrt{2^nn!\sqrt{\pi}l}}H_n\left(\frac{y}{l}+s_\perp p_1 l\right)e^{ip_1x},
\end{split}
\end{equation}
where $l=1/\sqrt{eB}$ is the magnetic length. $H_n(x)$ is the Hermite polynomials, and $s_{\perp}\equiv \text{sign}(eB)$. The wave functions satisfy the conditions of normalizability and completeness. Because u and d quarks carry different electric charges, the polarization functions acquire nontrivial Schwinger phases. Upon projection onto charged-meson Landau eigenstates, these phases combine with the external wave functions to produce gauge-invariant Landau-projected kernels. 
\begin{eqnarray}
\Pi_{ab,n}(q_0,B)=\int d^4xd^4y \varphi_{n}^*(x)\Pi_{ab}(x,y;B)\varphi_{n}(y).
\end{eqnarray}
Landau quantization therefore enters both through internal quark propagators and external meson legs, and their combined structure determines the magnetic-field dependence of the meson spectrum. Then the boson quadratic action \eqref{quadraaction} becomes a sum over Landau levels:
\begin{eqnarray}
S^{(2)}_\text{eff}
=\frac{1}{2}\sum_n\int\frac{dq_0}{2\pi}
\Phi_n^\dagger(q_0)
\mathcal K(q_0;B,n)
\Phi_n(q_0),
\end{eqnarray}
with $\Phi_n=(\pi_n,\rho_n)^T$ is the subspace of $\pi^+$ and $\rho^+_{s_z=0}$ at given Landau level $n$. Projecting the quadratic bosonic action onto mesonic Landau eigenstates yields the microscopic Landau-level projected inverse propagator,
\begin{eqnarray}
\label{microscopickernel}
&&\mathcal{K}(q_0;B,n)=\left(\begin{array}{cc}
\mathcal{K}_{\pi\pi}(q_0;B,n)  &  \mathcal{K}_{\pi\rho}(q_0,B,n) \\
 \mathcal{K}_{\rho\pi}(q_0;B,n)  &    \mathcal{K}_{\rho\rho}(q_0;B,n) \\
\end{array}\right),
\end{eqnarray}
with $\mathcal{K}_{\pi\pi}=\frac{1}{2G_S}-\Pi_{\pi\pi,n}(q_0,B)$, $\mathcal{K}_{\rho\rho}=\frac{1}{2G_V}-\Pi_{\rho\rho,n}(q_0,B)$, $\mathcal{K}_{\rho\pi}=-{\Pi}^3_{\rho\pi,n}(q_0,B)+\mathcal{K}_{\rho\pi}^{\text{tree}}(q_0,B)$, $\mathcal{K}_{\pi\rho}=-{\Pi}^3_{\pi\rho,n}(q_0,B)+\mathcal{K}_{\pi\rho}^{\text{tree}}(q_0,B)$. We focus on the case where the external pion and rho are in the lowest Landau level, namely take $n=0$ in the above expressions. After taking all the integrals, we finally arrive at 
\begin{equation}
\begin{split}
\label{polarization}
\Pi_{\pi\pi,0}&=-\frac{4N_c eB}{3\pi}\int\frac{dk_3}{(2\pi)}\sum_{i,j=0}^\infty \Big[Y_1(i,j) I_3(q_0)
\nonumber\\
&+\Big((k_3^2+m^2) Y_1(i,j)+\frac{8}{3l^2}Y_2(i,j)\Big)I_2(q_0)\Big],\nonumber\\
\Pi_{\rho\rho,0}&=-\frac{4N_c eB}{3\pi}\int\frac{dk_3}{(2\pi)}\sum_{i,j=0}^\infty \Big[Y_1(i,j) I_3(q_0)\nonumber\\
&+\Big((-k_3^2+m^2) Y_1(i,j)+\frac{8}{3l^2}Y_2(i,j)\Big)I_2(q_0)\Big],\nonumber\\
\Pi_{\rho\pi,0}^3&\equiv~-iq_0\;eB\;g_{\rho\pi}^\text{loop},\nonumber\\
\Pi_{\pi\rho,0}^3&\equiv~+iq_0\;eB\;g_{\rho\pi}^\text{loop},
\end{split}
\end{equation}
with 
\begin{eqnarray}
\label{gloop}
g_{\rho\pi}^\text{loop}&=&-\frac{4N_c}{3\pi}m_f\int\frac{dk_3}{(2\pi)}\sum_{i,j=0}^\infty Y_3(n,m) I_2(q_0).
\end{eqnarray}
The functions $I_2(q_0)$ and $I_3(q_0)$ encode the Matsubara sum, while coefficients $Y_i$ originate from the integral over transverse momentum, they are presented in Appendix.\ref{App1}.

In the numerical process, we use the Pauli Villars regulation with three regulators $a_i=\{0,1,2,3\}$, $c_i=\{1,-3, 3,-1\}$. The model parameters are chosen to be $\Lambda=1.25$GeV, $G_S=3.1$GeV$^{-2}$, $G_V=5.019$GeV$^{-2}$ for all schemes except the bosonization. These parameters correspond to $m_\pi=0.133$GeV, $m_\rho=0.77$GeV. The summation of Landau level is performed to over 200 Landau levels, which is sufficient for the range of magnetic field considered.  

\subsection{Tree-level $\pi-\rho$ mixing operator and vacuum matching}
\label{matching}
In addition to the loop-induced mixing generated by the quark polarization tensor, we include a gauge-invariant tree-level operator,
\begin{eqnarray}
\label{counterterm}
\mathcal L_{\text{tree}}
=\kappa_\text{tree}\Big(
\rho^+_\mu\tilde F^{\mu\nu} D_\nu \pi^- +
\rho^-_\mu\tilde F^{\mu\nu} D_\nu \pi^+
\Big),
\end{eqnarray}
whose coefficient $\kappa_\text{tree}$ has mass dimension $-1$. This operator is the lowest-order local term that is linear in the external electromagnetic field and consistent with gauge invariance, and it provides a direct $\pi-\rho$ transition in the effective meson theory.

In the present framework, the loop calculation gives the part of the $\rho\pi\gamma$ vertex that is explicitly generated by the quark polarization function. However, that loop contribution does not saturate the full physical vacuum coupling inferred from the radiative decay $\rho^\pm\rightarrow\pi^\pm\gamma$. The role of $\mathcal L_{\text{tree}}$ is therefore not to introduce an unrelated new mechanism, but to parametrize the remaining local short-distance contribution that is not explicitly resolved in the present microscopic calculation. In this sense, the tree term should be viewed as an effective counterterm fixed by matching to the physical vacuum transition.

It is therefore useful to isolate the loop-induced mixing strength coming from the quark polarization function in the weak-field limit. This has the same origin as the triangle diagram in the vacuum. In practice, we expand the off-diagonal polarization function to linear order in the external magnetic field and extract the coefficient of the structure $iq_0 eB$, which defines the vacuum loop contribution $g_{\rho\pi}^\text{loop}$. For a quark of flavor $f$ and charge $q_f=Q_f e$, in a constant magnetic field along z, the translationary invariant part of the propagator can be expanded as 
\begin{eqnarray}
S_f(k;B)=S_f^{(0)}(k)+S_f^{(1)}(k;B)+O(B^2)
\end{eqnarray}
with 
\begin{equation}
\begin{split}
S_f^{(0)}(k)&=\frac{k\cdot\gamma+M}{k^2-M^2+i\epsilon},\\
S_f^{(1)}(k;B)&=i q_f B\frac{\gamma^1\gamma^2(k_\parallel\cdot \gamma_\parallel+M)}{(k^2-M^2+i\epsilon)^2}
\end{split}
\end{equation}
here $k_\parallel^\mu=(k^0,0,0,k^3)$, $k_\perp^\mu=(0,k^1,k^2,0)$. To the first order of magnetic field, the mixing polarization function becomes 
\begin{equation}
\begin{split}
\Pi_{\rho_+\pi_+}^{3}
=&-i2N_c\int\frac{d^4k}{(2\pi)^4}\text{Tr}\Big[\gamma^3{S}_u^{(0)}(k) i\gamma_5 {S}_d^{(1)}(p)\\
&\qquad\qquad\qquad\quad~~ +\gamma^3{S}_u^{(1)}(k) i\gamma_5 {S}_d^{(0)}(p)\Big]\\
=&-iq_0 eB g^\text{loop,vac}_{\rho\pi},
\end{split}
\end{equation}
with
\begin{equation}
\begin{split}
g^\text{loop,vac}_{\rho\pi}
&=\frac{8}{3}N_c m_f\int \frac{d^3k}{(2\pi)^3}\Big[-\frac{n'_F(E_f)}{2E_f^2 (q_0^2-4 E_f^2)}\\
&\qquad~~ +\frac{(12 E_f^2-q_0^2)}{4 E_f^3 (q_0^2-4 E_f^2)^2}(1-2 n_F(E_f))\Big]. 
\end{split}
\end{equation}
Inserting the corresponding parameters, namely, in the vacuum $m_f=0.4$GeV, and at the pion pole mass $q_0=0.135$GeV. This gives $g^\text{loop,vac}_{\rho\pi}=0.064556$.

The off-diagonal polarization function in a background magnetic field and the vacuum $\rho\pi\gamma$ transition vertex are not independent mechanisms. Rather, they originate from the same underlying quark-level triangle diagram. In the weak-field expansion, the magnetic field enters through a single electromagnetic insertion on the quark line, and the resulting $\rho-\pi$ mixing amplitude has the same microscopic origin as the quark triangle diagram for the radiative transition $\rho\rightarrow \pi\gamma$. In this sense, the linear-B polarization coefficient is viewed as the soft external-field limit of the same $\rho\pi\gamma$ vertex that controls the vacuum radiative decay. Therefore, the weak-field polarization calculation provides the natural microscopic loop contribution to the effective $\pi-\rho$ mixing kernel.

At the same time, the weak-field loop contribution obtained in the present truncation does not fully saturate the effective vacuum coupling inferred from the physical decay width. This observation is central for the matching procedure adopted below. Instead of identifying the loop result with the full physical $\rho\pi\gamma$ coupling, we regard it as the explicitly calculable microscopic part, $g^\text{loop,vac}_{\rho\pi}$, and determine the remaining contribution phenomenologically from vacuum matching. 

From the decay process $\rho(P,\varepsilon_\rho)\rightarrow\pi(p)+\gamma(q,\varepsilon_\gamma)$ with $P=p+q$, the tree-level vertex \eqref{counterterm} gives the amplitude 
\begin{eqnarray}
i\mathcal{M}
&=&i\kappa_\text{exp} \varepsilon_\rho^\mu \tilde{F}_{\mu\nu}(-ip^\nu)
=i\kappa_\text{exp}\epsilon_{\mu\nu\alpha\beta}\varepsilon_\rho^\mu\varepsilon_{\gamma}^\beta p^\nu q^\alpha.
\end{eqnarray}
The spin-averaged squared amplitude is then $\overline{|\mathcal{M}|^2}=\frac{1}{3}\sum_{\text{pol}. \rho,\gamma}|\mathcal{M}|^2=\overline{|\mathcal{M}|^2}=\frac{2}{3}\kappa_\text{exp}^2(P\cdot q)^2$. And in the $\rho$ rest frame this leads to the standard two-body decay width
\begin{eqnarray}
\Gamma(\rho\rightarrow\pi\gamma)=\frac{|\vec{q}|}{8\pi m_\rho^2}\overline{|\mathcal{M}|^2}=\frac{\kappa_\text{exp}^2}{96\pi m_\rho^3}(m_\rho^2-m_\pi^2)^3.
\end{eqnarray}
Using the experimental partial width $\Gamma_{\text{exp}}(\rho^\pm\rightarrow\pi^\pm\gamma)\approx 0.068$ MeV, one obtains 
\begin{eqnarray}
\kappa_\text{exp}\approx 0.222~\text{GeV}^{-1}. 
\end{eqnarray}
In order to express the mixing strength in terms of canonically normalized mesonic fields, one must specify the vacuum wave-function renormalization factors of the pion and rho channels,
\begin{eqnarray}
\label{vacuum_Z}
Z_\pi^\text{vac}=\frac{\partial \Pi_{\pi\pi}(q)}{\partial q_0^2}\Big|_{q_0^2=m_\pi^2}, \qquad Z_\rho^\text{vac}=\frac{\partial \Pi_{\rho\rho}(q)}{\partial q_0^2}\Big|_{q_0^2=m_\rho^2}.
\end{eqnarray}
With this convention, the pole residue is given by $(Z_\phi^\text{vac})^{-1}$. 
This gives $Z_\pi(0)=0.109$ and $Z_\rho(0)=0.413$. With this convention, the loop and tree contributions combine to reproduce the physical vacuum coupling,
\begin{eqnarray}
\frac{g^\text{tree}_{\rho\pi}+g^\text{loop,vac}_{\rho\pi}}{\sqrt{Z_\pi^\text{vac}Z_\rho^\text{vac}}}=\frac{1}{e}\kappa_\text{exp}\approx 0.73337\;\text{GeV}^{-1},
\end{eqnarray}
where $e=\sqrt{4\pi\alpha}$. This matching condition fixes the local tree-level contribution $g^\text{tree}_{\rho\pi}= 0.09104\;\text{GeV}^{-1}$ once the vacuum loop part $g^\text{loop,vac}_{\rho\pi}$ is known.

It is useful to stress once more that $g^\text{tree}_{\rho\pi}$ should not be interpreted as a fundamental bare coupling independent of the loop term. Rather, within the present effective description it represents the part of the physical $\rho\pi\gamma$ vertex that is not explicitly generated by the microscopic loop calculation. The total effective mixing is therefore
\begin{eqnarray}
g_{\rho\pi}(B)=g^\text{loop}_{\rho\pi}(B)+g^\text{tree}_{\rho\pi},
\end{eqnarray}
where the magnetic-field dependence of the loop part is taken from the explicit polarization calculation, while the tree part is fixed from the vacuum radiative decay and then kept as a local effective contribution.

\subsection{Near-pole expansion}
\label{nearpole}
The low-energy dynamics relevant for lattice observables is governed by the lowest Landau level. We therefore construct a near-pole effective description for the LLL by expanding the diagonal elements of the Landau-projected kernel \eqref{microscopickernel} around their respective poles and canonically normalizing the fields,
\begin{eqnarray}
K_{\phi\phi}\simeq Z_\phi(B)(q_0^2-E_\phi^2(B)), \quad \phi=\pi,\rho.
\end{eqnarray}
$E_\pi(B)$ and $E_\rho(B)$ are are the unmixed lowest-Landau-level energies of the pion and longitudinal rho meson, respectively. The wave-function renormalizations $Z_\pi(B)$ and $Z_\rho(B)$ at the pole encode the sensitivity of each mode to the magnetic background. This procedure isolates the relevant collective modes and makes the dynamical role of residues and mixing terms transparent. 

After canonical normalization, the LLL dynamics in the closed subspace of $\pi^+$ and longitudinal $\rho^+_{s_z=0}$ is captured by the near-pole quadratic kernel
\begin{eqnarray}
\label{determinantK}
\mathcal{K}_c(q_0)\approx\left(\begin{array}{cc}
q_0^2-E_\pi^2(B)  &  -i q_0 h(B) \\
i q_0 h(B) &    q_0^2-E_\rho^2(B) \\
\end{array}\right).
\end{eqnarray}
The off-diagonal terms describe magnetic field induced mixing between the pion and the longitudinal rho mode. Their strength is parametrized as
\begin{eqnarray}
\label{hB}
h(B)=\frac{eB[g^\text{loop}_{\rho\pi}(B)+g^\text{tree}_{\rho\pi}]}{\sqrt{Z_\pi(B) Z_\rho(B)}},
\end{eqnarray}
where $g^\text{loop}_{\rho\pi}(B)$ arises from quark-loop polarization in the magnetic field, while $g^\text{tree}_{\rho\pi}$ denotes the local coupling fixed by the vacuum decay $\rho^\pm\rightarrow\pi^\pm\gamma$. The physical eigenmodes within this near-pole effective theory are obtained from the pole condition
\begin{eqnarray}
\label{polecondition}
\text{det}\;\mathcal{K}_c(q_0)=0. 
\end{eqnarray}

The structure of the kernel makes the physical mechanism transparent: as the magnetic field increases, the mixing induces level repulsion between a light pion mode $\pi^+$ and a heavier longitudinal rho mode $\rho^+_{s_z=0}$. When the residue $Z(B)$ decreases rapidly, the effective mixing strength $h(B)$ is strongly amplified near the pole, leading to a downward bending of the lower eigenmode. This residue-enhanced level repulsion provides a natural route to non-monotonic magnetic-field dependence of the charged pion energy.

It is worth noticing that this procedure assumes that the relevant collective mode remains adiabatically connected to the unmixed LLL pole, an assumption whose validity can be tested by comparison directly solving the determinant of the full Landau-projected kernel \eqref{microscopickernel}.

\section{Numerical results}
\label{result}
\begin{figure}[t]
\centering
\includegraphics[width=0.4\textwidth]{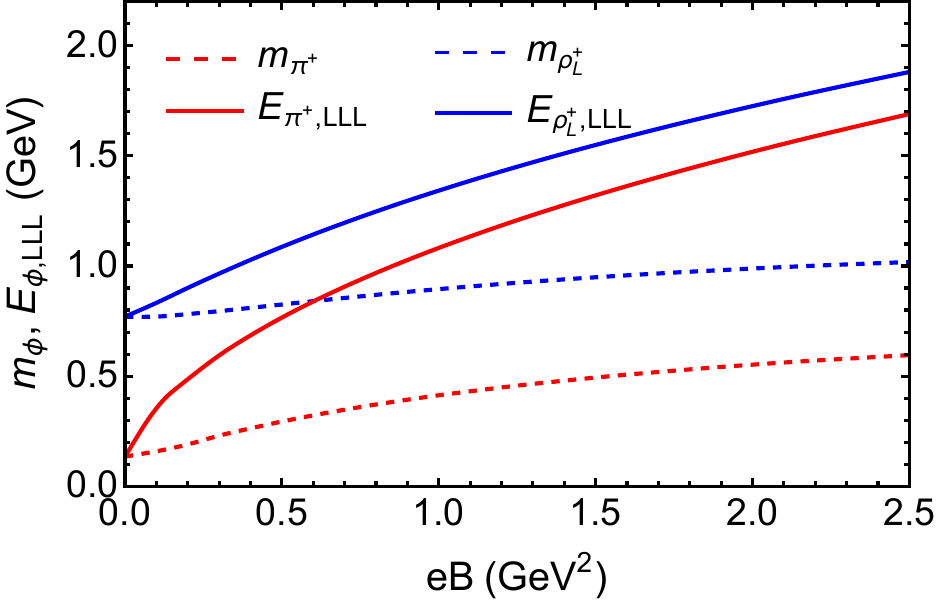}
\caption{Unmixed lowest-Landau-level (LLL) energies $E_\pi(B)$ and $E_\rho(B)$ obtained from the diagonal Landau-projected kernels \eqref{microscopickernel}, together with the corresponding rest masses $m_\phi=\sqrt{E_\phi^2-eB}$. Both unmixed LLL energies increase monotonically with the magnetic field.}
\label{fig1}
\end{figure}

We first determine the unmixed LLL energies of the charged pion and longitudinal rho meson from the diagonal Landau-projected kernels. As shown in Fig.\ref{fig1}, both $E_\pi(B)$ and $E_\rho(B)$ increase monotonically with magnetic field and exhibit the expected $\sqrt{eB}$ behavior at large fields, reflecting Landau quantization. When converted to rest masses via $m_\phi=\sqrt{E_\phi^2-eB}$, the charged pion mass remains monotonic. The strict monotonicity of the unmixed LLL energies demonstrates that Landau quantization and magnetic catalysis alone cannot generate a turnover.

\begin{figure}[t]
\centering
\includegraphics[width=0.4\textwidth]{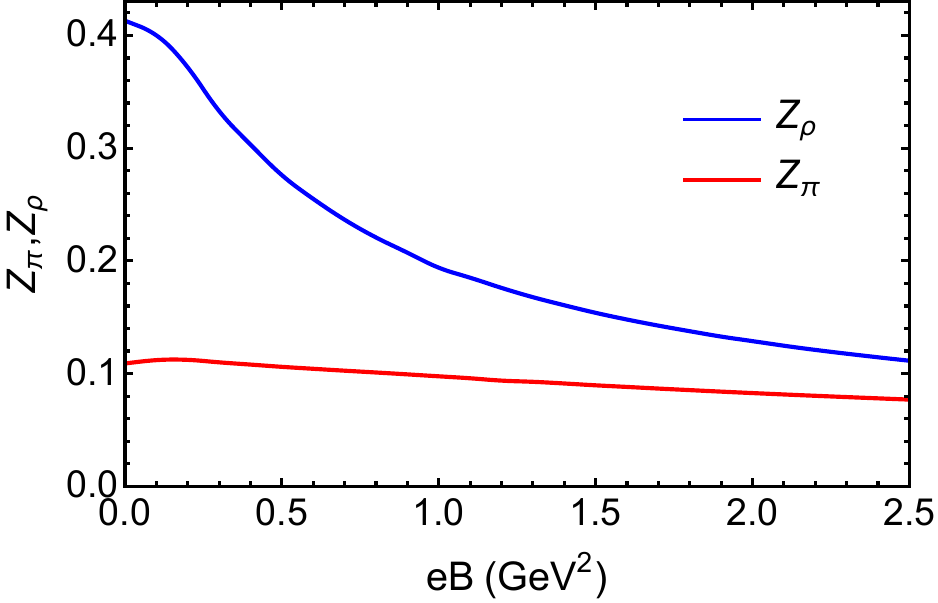}
\caption{Wave-function renormalizations $Z_\pi(B)$ and $Z_\rho(B)$ extracted from the slopes of the diagonal LLL Landau-projected kernels at the poles. While $Z_\pi(B)$  decreases mildly with increasing magnetic field, $Z_\rho(B)$ is rapidly suppressed in the LLL.}
\label{fig2}
\end{figure}

The wave-function renormalizations extracted from the near-pole slopes are shown in Fig.\ref{fig2}. A crucial feature is the qualitatively different magnetic-field dependence in the two channels. While $Z_\pi(B)$ decreases mildly, $Z_\rho(B)$ is rapidly suppressed in the LLL and becomes very small at strong fields. Physically, the suppression of $Z_\rho(B)$ reflects the restructuring of the longitudinal vector correlator in a magnetic background. The collapse of $Z_\rho(B)$ dynamically amplifies mixing effects near the pole.

\begin{figure}[t]
\centering
\includegraphics[width=0.4\textwidth]{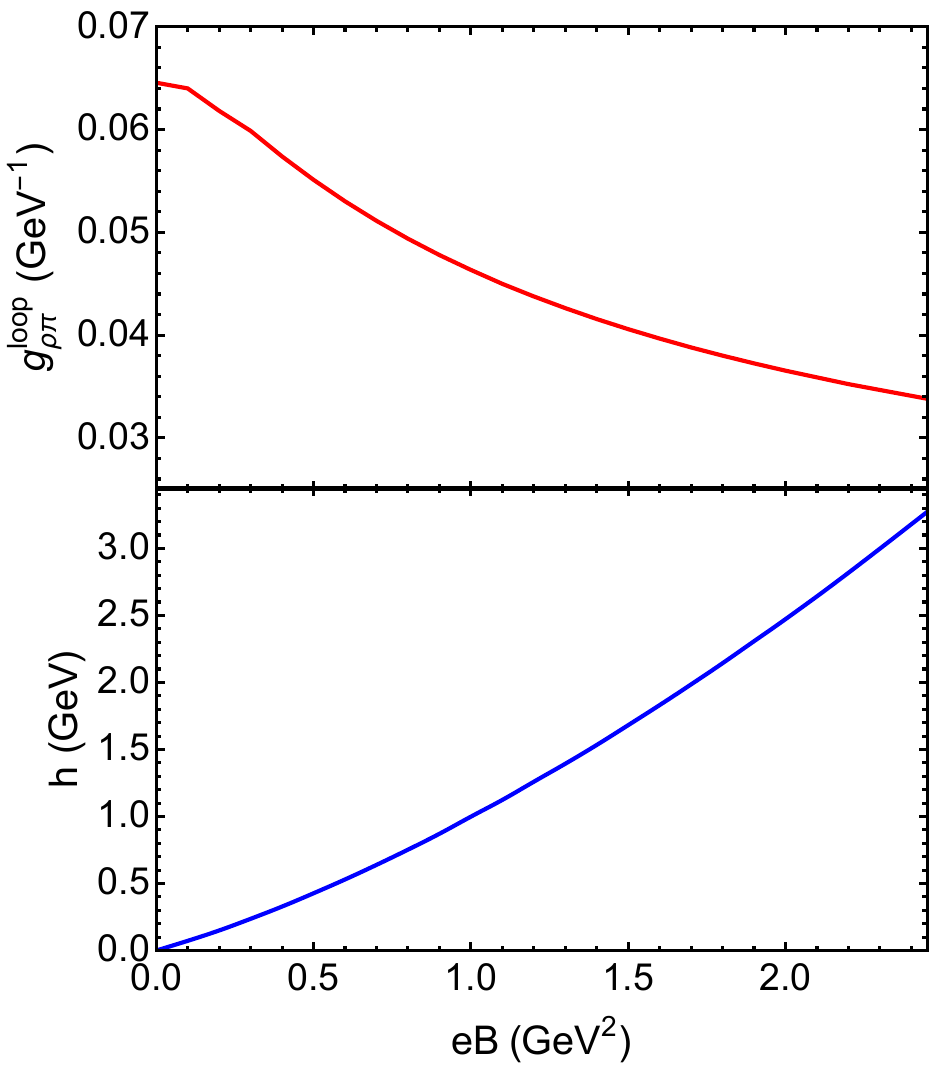}
\caption{
Upper panel: Loop-induced mixing coupling $g_{\rho\pi}^\text{loop}(B)$ defined in \eqref{gloop}, evaluated at $q_0 = E_\pi(B)$. Lower panel: Total mixing strength $h(B)$ defined in \eqref{hB}, including both loop-induced and tree-level contributions.}
\label{fig3}
\end{figure}

The loop-induced mixing coupling $g_{\rho\pi}^{\rm loop}(B)$, defined in \eqref{gloop}, is shown in the upper panel of Fig.\ref{fig3} and evaluated at $q_0=E_\pi(B)$. The loop contribution encodes infrared quark dynamics in a magnetic field and is smaller than the tree-level term fixed by the vacuum decay $\rho^\pm\to \pi^\pm+\gamma$, as expected in an EFT description. The total mixing strength $h(B)$, shown in the lower panel, grows strongly with increasing magnetic field. Although $g_{\rho\pi}^{\rm loop}(B)$ alone decreases with $B$, the total mixing is enhanced by the explicit $eB$ factor and, more importantly, by the suppression of $Z_\rho(B)$ in the denominator. The effective near-pole mixing is therefore dynamically amplified even when the microscopic loop contribution remains moderate.

\begin{figure}[t]
\centering
\includegraphics[width=0.4\textwidth]{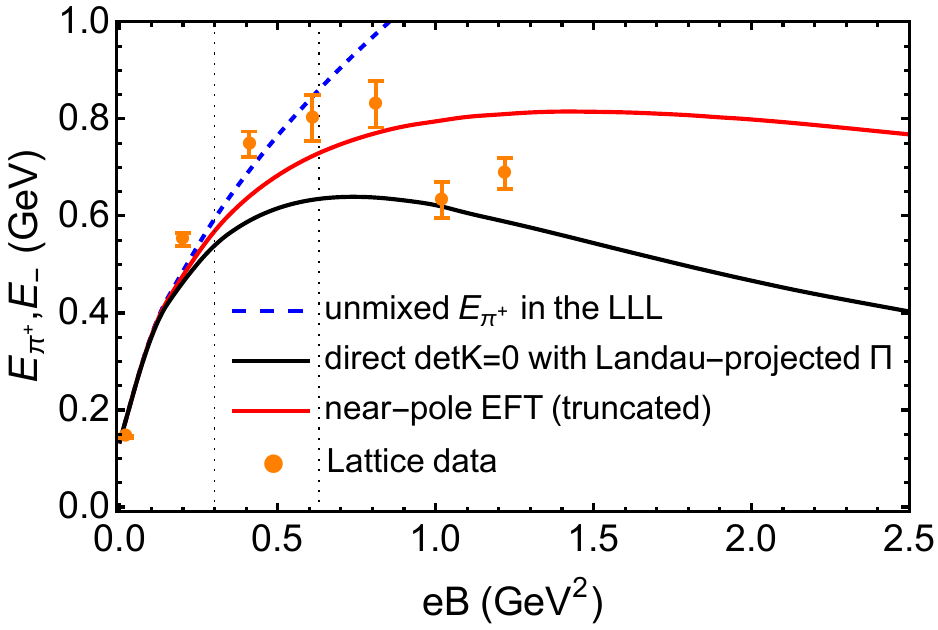}
\caption{Lowest eigenmode of the coupled $\pi^+-\rho^+_{s_z=0}$ system in the LLL as a function of the magnetic field. The dashed blue curve shows the unmixed pion LLL energy $E_{\pi^+}$. The red curve corresponds to the lower eigenmode $E_-$ obtained from the near-pole effective kernel in Eq.~\eqref{determinantK}, while the black curve is obtained by solving $\det K(q_0)=0$ directly using the Landau-level projected kernel in Eq.~\eqref{microscopickernel}, without invoking the near-pole expansion. The orange points with error bars are the continuum-extrapolated charged-pion lattice data from Ref.~\cite{Ding:2026qzu}, shown for comparison. The vertical dotted lines indicate the characteristic field range where the lattice data begin to deviate from the simple LLL trend and where the turnover reaches its maximum.}
\label{fig4}
\end{figure}

Solving the pole condition \eqref{polecondition} yields two mixed eigenmodes. The lower eigenmode $E_-$ is shown in Fig.\ref{fig4}. At weak fields it closely follows the unmixed pion LLL energy. At intermediate fields, increasing mixing induces level repulsion between the light pion mode and the heavier longitudinal rho mode, causing the lower eigenvalue to bend downward. This turnover is a direct consequence of residue-enhanced pion-rho mixing within the same Landau level. For comparison, Fig.\ref{fig4} also shows the result obtained by solving $\text{det}\,\mathcal{K}(q_0)=0$ directly using the full Landau level projected kernel \eqref{microscopickernel}. At weak and intermediate fields the two approaches agree, confirming that the near-pole formulation captures the essential determinant physics. At larger $eB$ the increasing deviation reflects sensitivity to higher-order energy dependence and provides a quantitative estimate of systematic uncertainties in the truncated near-pole description. 

Lattice simulations extract an effective mass from the long-time behavior of charged pion correlators, corresponding to the lowest physical eigenmode rather than to the unmixed pion Landau level. The lattice continuum-extrapolated data are included in Fig.~\ref{fig4} for direct comparison. One sees immediately that the unmixed LLL energy fails to account for the observed non-monotonic behavior, whereas both the near-pole and direct-determinant treatments generate a turnover. This supports the underlying $\pi-\rho$ mixing mechanism itself. At the same time, the comparison also reveals a quantitative difference between the two extraction strategies. The near-pole truncation places the maximum at larger $eB$ and yields a relatively mild downward bending, while the direct determinant solution shifts the turnover toward the lattice-indicated region around $eB \sim 0.6$--$0.7~\mathrm{GeV}^2$. In this sense, the full determinant treatment provides a closer description of the lattice trend. The remaining discrepancy at larger $eB$ indicates that, although the mechanism is robust, the precise location and shape of the turnover remain sensitive to the detailed structure of the kernel and to the extraction procedure.

\section{Summary and discussion}
\label{summary}
We have studied magnetic field induced $\pi-\rho$ mixing as a mechanism for the non-monotonic behavior of the charged pion mass in a magnetic field. Within a near-pole effective description derived from the Landau projected NJL kernel, we find that the LLL pion mixes with the longitudinal rho mode, and that this level repulsion is strongly amplified by the suppression of the rho residue near the pole. The resulting lowest mixed mode develops a turnover with increasing magnetic field.

Comparison with the direct determinant solution shows that the same mechanism persists beyond the truncated near-pole description, although the quantitative location of the maximum is sensitive to the detailed structure of the kernel. When compared with the lattice continuum-extrapolated charged pion data, both treatments reproduce the qualitative non-monotonic trend, while the direct determinant solution places the turnover closer to the lattice-indicated region than the near-pole truncation. Our results therefore support residue-enhanced $\pi-\rho$ mixing as an important candidate mechanism for the charged-meson behavior seen in lattice QCD.

More broadly, the present analysis highlights that residue effects can qualitatively reshape charged meson spectra in strong magnetic fields. Possible extensions include mixing with axial-vector modes in the light sector and analogous effects in strange meson systems. A more systematic comparison among different extraction schemes and matching prescriptions will be presented elsewhere.

\acknowledgments
The author would like to thank Pengfei Zhuang, Shijun Mao, Hengtong Ding for fruitful discussion. The work is supported by NSFC grant Nos. 12005112 and The Fundamental Research Funds for Beijing Municipal Universities.

\appendix
\section{Landau projected polarization function}
\label{App1}
The functions $I_2(q_0)$ and $I_3(q_0)$ encodes the Matsubara sum, 
\begin{equation}
\begin{split}
I_2
&=\frac{1}{4E_{u,i}E_{d,j}}\Big(\frac{n_F(E_{d,j})-n_F(E_{u,i})}{(E_{d,j}-E_{u,i}+q_0)}
+\frac{n_F(E_{d,j})-n_F(E_{u,i})}{(E_{d,j}-E_{u,i}-q_0)}\nonumber\\
&+\frac{1-n_F(E_{d,j})-n_F(E_{u,i})}{(E_{d,j}+E_{u,i}-q_0)}
+\frac{1-n_F(E_{d,j})-n_F(E_{u,i})}{(E_{d,j}+E_{u,i}+q_0)}\Big),\nonumber\\
I_3&=\frac{1}{4}\Big(\frac{n_F(E_{d,j})-n_F(E_{u,i})}{(-E_{d,j}+E_{u,i}-q_0)}
+\frac{n_F(E_{d,j})-n_F(E_{u,i})}{(-E_{d,j}+E_{u,i}+q_0)}\nonumber\\
&+\frac{1-n_F(E_{d,j})-n_F(E_{u,i})}{(E_{d,j}+E_{u,i}-q_0)}
+\frac{1-n_F(E_{d,j})-n_F(E_{u,i})}{(E_{d,j}+E_{u,i}+q_0)}\Big),
\end{split}
\end{equation}
with $E_{u,i}^2=2i|q_uB|+k_3^2+m_f^2$ and $E_{d,j}^2=2j|q_dB|+k_3^2+m_f^2$. The coefficients $Y_i$ are defined as
\begin{equation}
\begin{split}
Y_1(m,n)&=-\frac{2^{m-1}}{3^{n+m}}\frac{(m+2n)\Gamma(n+m)}{\Gamma(1+n)\Gamma(1+m)},\nonumber\\
Y_2(m,n)&=-\frac{ 2^{m-1}}{3^{n+m}}\frac{\Gamma(n+m)}{\Gamma(n)\Gamma(m)},\nonumber\\
Y_3(n,m)&=-\frac{2^{m-1}}{3^{n+m}}\frac{(m-2n)\Gamma(n+m)}{\Gamma(1+n)\Gamma(1+m)}.
\end{split}
\end{equation}

\bibliographystyle{unsrt} 
\bibliography{ref}

\end{document}